\documentclass[amsmath, amsfonts, superscriptaddress, showpacs, twocolumn, prb]{revtex4}
\usepackage{graphicx}
\usepackage{epsfig}
\usepackage{bm}
\usepackage{dcolumn}
\usepackage{amsmath}
\usepackage{amssymb}

\newcommand{\dv}{\mathop{\rm div}\nolimits}

\begin{document}

\title{Hydrodynamic Coulomb drag of strongly correlated electron
liquids}

\author{S.~S.~Apostolov}
\affiliation{Department of Physics and Astronomy, Michigan State
University, East Lansing, Michigan 48824, USA}

\author{A.~Levchenko}
\affiliation{Department of Physics and Astronomy, Michigan State
University, East Lansing, Michigan 48824, USA}

\author{A.~V.~Andreev}
\affiliation{Department of Physics, University of Washington,
Seattle, Washington 98195, USA}

\begin{abstract}
We develop a theory of Coulomb drag in ultraclean double layers with
strongly correlated carriers. In the regime where the equilibration
length of the electron liquid is shorter than the interlayer spacing
the main contribution to the Coulomb drag arises from hydrodynamic
density fluctuations. The latter consist of plasmons driven by
fluctuating longitudinal stresses, and diffusive modes caused by
temperature fluctuations and thermal expansion of the electron
liquid. We express the drag resistivity in terms of the kinetic
coefficients of the electron fluid. Our results are nonperturbative
in interaction strength and do not assume Fermi-liquid behavior of
the electron liquid.
\end{abstract}

\date{\today}

\pacs{71.27.+a, 72.10.-d, 73.40.Ei, 73.63.Hs}

\maketitle

\textit{Introduction and motivation}. Interaction-induced mutual
friction phenomena in the electrically disconnected double quantum
wells provide a uniquely sensitive probe of electronic scattering
and correlations. The effect, commonly referred to as Coulomb drag
(CD),~\cite{Pogrebinskii,Price} consists of passing a drive current
$I$ through one (active) layer and measuring nonlocally induced drag
voltage $V_D$ in the other (passive) layer. In experiments, the drag
resistivity $\rho_D=V_D/I$ is usually studied as a function of
temperature, magnetic field, electron density, and interlayer
separation.~\cite{Solomon,Sivan,Gramila,Lilly,Kellogg,Pillarisetty}
Recent measurements of $r_D$ in the tightly nested graphene double
layers~\cite{Tutuc,Geim} triggered resurgence of interest and new
proposals for the mechanisms of CD effect.~\cite{Levitov,Mirlin}

The magnitude and even the mechanism of CD depend on the temperature
$T$, interlayer distance $d$, properties of the disorder potential,
and the strength of electron-electron interactions. The latter is
characterized by the interaction parameter $r_s=(\pi n
a_B^2)^{-1/2}$, where $n$ is the electron density and
$a_B=\epsilon/m e^2 $ is the effective Bohr radius in the material
($\epsilon$ being the dielectric constant, hereafter $\hbar=1$).

In the weakly interacting regime, $r_s\lesssim1$, measurements of
$\rho_D(T)$~\cite{Solomon,Gramila,Sivan} are in qualitative
agreement with the predictions of the Fermi-liquid
theory.~\cite{Laikhtman,Smith,MacDonald,Kamenev,Flensberg}
Specifically, the drag resistivity is relatively small,
$\rho_D/\rho_Q\propto(k_Fd)^{-\alpha_d}(T/E_F)^{\alpha_T}$, scales
quadratically with the temperature at $T\ll E_F$ ($\alpha_T=2$ up to
logarithmic corrections in the disordered
case~\cite{MacDonald,Kamenev}), and is inversely proportional to a
certain power of interlayer distance $d$ ($\alpha_d=2-4$ depending
on the ratio between $d$ and electronic mean free path), where $k_F$
and $E_F$ are Fermi momentum and energy, respectively, and
$\rho_Q=2\pi/e^2$ is the resistance quantum. At higher temperatures,
$T\sim E_F$, drag is enhanced by the plasmons such that $\rho_D/T^2$
has a relatively broad peak structure near the characteristic energy
of the plasmon modes.~\cite{Flensberg-plasmons} There are few
exceptions where the CD effect was studied beyond the leading order
in interlayer interaction. It was shown in Ref.~\onlinecite{AL} that
interference corrections to $\rho_D$ originating from the
third-order processes in the interlayer interaction, are
increasingly important at low $T$. This is in loose qualitative
agreement with the fact that experimental values of the drag are
larger than conventional values that vanish as $T\to0$, however, the
theory of Ref.~\onlinecite{AL} still assumes relatively weak
interactions.

For $r_s\gg1$ Coulomb drag is not well understood. In this case
apart from the Fermi energy $E_F$ there are two other important
energy scales: the interaction energy $V=r_sE_F$ and the Debye
frequency $\Omega_D\sim\sqrt{VE_F}=\sqrt{r_s}E_F$. As a consequence
of this hierarchy $E_F\ll\Omega_D\ll V$ there is a wide temperature
interval, $E_F< T< V$, in which the system remains strongly
correlated but the Fermi-liquid description does not apply. It is
worth noting that such strongly correlated liquids may be treated
classically only for $T>\Omega_D$, while at lower temperatures they
form a semiquantum state.~\cite{AFAndreev,Review,Note} Microscopic
theory of electron transport in this very interesting regime has not
been developed.

In samples with $r_s\gg 1$ even at low temperatures, $T\ll E_F$, the
drag resistance  is one to two orders of magnitude larger than
expected on the basis of a simple extrapolation of the small $r_s$
results.~\cite{Pillarisetty} Furthermore, the power exponent
$2<\alpha_T<3$ deviates from its nominal value and the system has
anomalous response in a magnetic field.~\cite{Pillarisetty} A
detailed study based on an extrapolation of Fermi-liquid-based
formulas to the region where $r_s\gtrsim1$ has been carried out in
Ref.~\onlinecite{Hwang} in an attempt to address the data of
Ref.~\onlinecite{Pillarisetty}. Finally two elegant phenomenological
theories designed for the strongly disordered electronic
systems~\cite{Raikh} and electronic microemulsions~\cite{Spivak}
have been proposed.

Most of the previous theoretical work on CD focused on the
collisionless regime, in which the spacing between the layers is
smaller than the mean free path of the quasi-particles. In this
paper we develop a theory of Coulomb drag in the opposite regime,
where the density fluctuations of the electron liquid responsible
for Coulomb drag may be described using the hydrodynamic approach. A
hydrodynamic theory of resistivity was recently formulated in
Ref.~\onlinecite{Andreev}. We generalize this theory to the case of
drag resistivity in double layer systems. This requires
consideration of the fluctuation corrections to
hydrodynamics.~\cite{LL,PL} We identify a mechanism of drag
resistivity originating from the entropy fluctuations, which due to
the thermal expansion changes the electron density thus leading to
the Coulomb coupling between the layers. The contribution from the
plasmon modes is also discussed in details.

We assume that the interlayer distance exceeds the equilibration
length of the electron liquid, $d\gg\ell_{\mathrm{ee}}$, and at the
same time the phonon contribution to drag is negligible.
Hydrodynamic theory of phonon-mediated drag for charge neutral
liquids was developed in Ref.~\onlinecite{Meierovich}. The
hydrodynamic description applies to any liquid type. Microscopic
properties of the liquid manifest themselves via the temperature and
density dependence of the kinetic coefficients. Experimentally, the
hydrodynamic regime is likely to be relevant to clean low carrier
density systems, e.g., hole systems of
Refs.~[\onlinecite{Pillarisetty,Gao}] with the typical
$r_s\sim10-40$ and Fermi energies on the order of Kelvin. In this
case, at $T\gtrsim E_F$, electron-phonon scattering is still
weak.~\cite{Gao} On the other hand,  carriers form a nondegenerate
strongly correlated liquid (semiquantum or classical), and the
hydrodynamic description applies from very short distances of the
order of inter-electron spacing. Drag resistivity measurements in a
high mobility quantum well with $r_s\sim1$ and at large interlayer
spacing ($d\sim 5000 {\AA}$) have been reported in the
literature~\cite{Eisenstein} and attributed to phonon
drag.~\cite{MacDonald}

\textit{Stochastic Navier-Stokes equations}. If the interlayer
spacing $d$ exceeds the equilibration length
$d\gg\ell_{\mathrm{ee}}$ Coulomb drag is dominated by hydrodynamic
density fluctuations.  The latter obey Gaussian distribution and may
be described by introducing stochastic Langevin forces into the
hydrodynamic equations.~\cite{LL} Having in mind linear response
theory we start from the linearized continuity, Navier-Stokes and
entropy production equations for a fluid moving in an external
potential $U$,
\begin{eqnarray}
&&\partial_t\delta n+\partial_{i}(\delta n v_i+n\delta
v_i)=0\label{Eq-cont},\\
&&mn(\partial_t+\mathbf{v}\cdot\nabla)\delta
v_i=- n \partial_i U-\partial_{i}\delta P+\partial_{k}\delta\sigma_{ik},\label{Eq-NS}\\
&&nT(\partial_t+\mathbf{v}\cdot\nabla)\delta
s=-\dv\mathbf{Q},\label{Eq-EP}
\end{eqnarray}
where $i,k$ are Cartesian indices and we used shorthand notation for
the spatial derivative $\partial_i=\partial/\partial x_i$. In
Eqs.~\eqref{Eq-cont}-\eqref{Eq-EP} $n$, $m$, and $T$ are,
respectively, the equilibrium particle density, mass, and
temperature; $\mathbf{v}$ is the uniform fluid velocity; while
$\delta n$, $\delta v_i$, $\delta s$, and $\delta P$ are
fluctuations of density, velocity, entropy per particle, and
pressure in the liquid flow. In a charged liquid, the external
potential $U$ is determined by the fluid density via the Poisson
equation. The set of these equations has to be replicated for both
active and passive layers and we will use subscripts
$\uparrow\downarrow$ to distinguish the two. The linearized viscous
stress tensor $\delta\sigma_{ik}$ and the thermal energy flux
$\mathbf{Q}$ are given, respectively, by
\begin{eqnarray}
&&\hskip-.5cm \delta\sigma_{ik}=\eta(\partial_k\delta
v_i+\partial_i\delta
v_k)+(\zeta-\eta)\delta_{ik}\dv\delta\mathbf{v}+\varsigma_{ik},\\
&&\hskip-.5cm \mathbf{Q}=-\kappa\nabla\delta T+\mathbf{g}.
\end{eqnarray}
Here $\kappa$ is the thermal conductivity, $\eta$ is the first
(shear) viscosity of the liquid, and $\zeta$ is the second (bulk)
viscosity. We consider a symmetric setup in which fluctuating
Langevin heat and stress fluxes in the two layers have identical
variances:
\begin{eqnarray}
&&\langle
\mathbf{g}_i(\mathbf{r},t)\mathbf{g}_j(\mathbf{r}',t')\rangle=2\kappa
T^2\delta_{ij}\delta(\mathbf{r}-\mathbf{r}')\delta(t-t'),\\
&&\langle\varsigma_{ik}(\mathbf{r},t)\varsigma_{lm}(\mathbf{r}',t')\rangle=
2T\delta(\mathbf{r}-\mathbf{r}')\delta(t-t')\nonumber\\
&&\times[\eta(\delta_{il}\delta_{km}+
\delta_{im}\delta_{kl})+(\zeta-\eta)\delta_{ik}\delta_{lm}],
\end{eqnarray}
where $\langle\ldots\rangle$ denotes averaging over the thermal
fluctuations. The steady current $\propto n\mathbf{v}$ in the active
layer exerts the drag force $\mathbf{F}_D=\langle\delta
n_\downarrow(-\nabla U_{\downarrow})\rangle$ on the passive layer.
Relating the potential to density fluctuations by using the Poisson
equation and ignoring the intralayer forces we can express the drag
force in terms of the density-density correlation function
\begin{equation}\label{F}
\mathbf{F}_D=\sum_{q,\omega}(-i\mathbf{q})\frac{2\pi e^2}{\epsilon
q}e^{-qd}\langle\delta n_\uparrow(\mathbf{q},\omega) \delta
n_\downarrow(-\mathbf{q},-\omega)\rangle
\end{equation}
where $\delta n_{\uparrow,\downarrow}(\mathbf{q},\omega)$ are the
Fourier components of the density fluctuations in both layers and
$q$ is the absolute value of the vector $\mathbf{q}$. Knowing the
drag force one readily finds the drag resistivity
$\rho_D=\mathbf{F}_D/\mathbf{v}e^2n^2$.

\textit{Results for the drag resistivity}. Our technical goal now is
to solve coupled equations \eqref{Eq-cont}-\eqref{Eq-EP} to the
linear order in $\mathbf{v}$. It will be convenient for our purposes
to choose entropy and density as independent variables, and thus
express temperature and pressure fluctuations via thermodynamic
relations. To this end, we rewrite
Eqs.~\eqref{Eq-cont}-\eqref{Eq-EP} in the Fourier components, relate
temperature fluctuations to the entropy and density $\delta
T=(\partial T/\partial s)_V\delta s+(\partial T/\partial n)_S\delta
n$, and exclude $\delta\mathbf{v}$ with the help of the continuity
equation \eqref{Eq-cont}. We thus find for the active layer
\begin{subequations}
\begin{equation}\label{Eq-rho}
(\nu q^2-i\omega+i\mathbf{qv})(\omega-\mathbf{qv})\delta
n_\uparrow=-\frac{iq^2}{m}(\delta
P_\uparrow+nU_\uparrow)+\frac{i\mathbf{q}}{m}(\hat{\varsigma}_\uparrow\mathbf{q}),
\end{equation}
\begin{equation}\label{Eq-S}
(\chi q^2-i\omega+i\mathbf{qv})\delta s_\uparrow+\chi q^2
\left(\frac{\partial s}{\partial n}\right)_T\!\!\!\delta
n_\uparrow=-\frac{i\mathbf{qg}_\uparrow}{nT}.
\end{equation}
\end{subequations}
Here $\nu=(\eta+\zeta)/mn$ is the sum of shear and bulk kinematic
viscosities and $\chi=\kappa/n c_v$ is the thermal diffusivity,
while $c_v=T(\partial s/\partial T)_V$ is the heat capacity. In the
passive layer we have the same set of equations but with
$\mathbf{v}=0$ and an interchange of indices
$\uparrow\to\downarrow$. It becomes apparent from the structure of
Eqs.~\eqref{Eq-rho} and \eqref{Eq-S} that even though we consider a
disorder-free system the entropy fluctuations in the liquid
propagate diffusively. Because of thermal expansion, they result in
diffusively spreading density fluctuations that occur at uniform
stress in the liquid. The pressure fluctuations $\delta P$ can be
expressed in terms of density and entropy in a given layer, while
the external potential $U$ is determined by the density fluctuations
in both layers. Specifically we have for the active layer
\begin{eqnarray}
\delta P_\uparrow+nU_\uparrow=\left(\frac{\partial P}{\partial
s}\right)_V\delta s_\uparrow+\left(\frac{\partial P}{\partial
n}\right)_S\delta n_\uparrow\nonumber\\
+\frac{2\pi n e^2}{\epsilon q}(\delta n_\uparrow+e^{-qd}\delta
n_\downarrow),\label{Eq-P}
\end{eqnarray}
whereas the pressure variation in the passive layer $\delta
P_\downarrow$ is obtained from above by interchanging indices
$\uparrow\rightleftarrows\downarrow$. At wavelengths longer than the
screening length the second term in the right-hand side of
Eq.~\eqref{Eq-P} is small in comparison to the third and can be
neglected. Physically this means that the dependence of the stress
on the density of the electron liquid is dominated by the long-range
Coulomb interaction. At the same time, the first term in the
right-hand side of Eq.~\eqref{Eq-P} must be retained because it
describes the dependence of the stress on the different
thermodynamic variable, $\delta s$. With the aid of Eq.~\eqref{Eq-P}
we can exclude entropy fluctuations from Eqs.~\eqref{Eq-rho} and
\eqref{Eq-S}, and thus arrive at two coupled linear algebraic
equations for the variances of thermally induced density
fluctuations between the layers
\begin{equation}\nonumber
\Pi_\pm\delta n_\pm=\frac{i\mathbf{qv}}{2}\left[\Gamma_+\delta
n_++\Gamma_-\delta n_--
\frac{\mathbf{q}}{m}(\hat{\varsigma}_+\mathbf{q})-
\frac{\mathbf{q}}{m}(\hat{\varsigma}_-\mathbf{q})\right]
\end{equation}
\begin{equation}
-\frac{iq^2}{m c_v}\left(\frac{\partial s}{\partial\ln n}\right)_T
(\mathbf{qg}_\pm)
-(\omega_\chi-i\omega)\frac{\mathbf{q}}{m}(\hat{\varsigma}_\pm\mathbf{q}),\label{Eq-rho-pm}
\end{equation}
where we introduced symmetric $(+)$ and antisymmetric $(-)$
combinations of the fields $\delta n_\pm=\delta n_\uparrow\pm\delta
n_\downarrow$, and similarly for all other quantities. The
propagator of the excitation modes $\Pi_\pm$ and the vertex function
$\Gamma_\pm$ are defined by the following expressions
\begin{subequations}
\begin{equation}
\Pi_\pm(\mathbf{q},\omega)=(\omega_\nu-i\omega)
(\omega_\chi-i\omega)i\omega-(\omega_\chi-i\omega)\omega^2_\pm-\omega_\chi\omega^2_\alpha,
\label{Pi} \end{equation}
\begin{equation}
\Gamma_\pm(\mathbf{q},\omega)=(\omega_\nu-2i\omega)(\omega_\chi-2i\omega)+(\omega^2+\omega^2_\pm).
\end{equation}
\end{subequations}
Here we introduced characteristic mode frequencies:
$\omega_{\chi}=\chi q^2$ and $\omega_{\nu}=\nu q^2$ correspond to
the thermal and viscous diffusion, the frequencies
$\omega^2_\pm=\omega^2_p(1\pm e^{-qd})$, with $\omega^2_p=2\pi
e^2nq/\epsilon m$, correspond to the plasmons, while
$\omega_\alpha=uq$ corresponds to the acoustic mode associated with
the thermal expansion of the fluid with the characteristic velocity
$u=\sqrt{T/mc_v}(\partial s/\partial\ln n)_T$. In deriving
Eq.~\eqref{Eq-rho-pm} we also made use of the thermodynamic relation
$(\partial P/\partial s)_V=n^2(\partial T/\partial n)_S$. We look
for the solution of Eq.~\eqref{Eq-rho-pm} to the linear order in
$\mathbf{v}$ in the form $\delta n_\pm=\delta n^{(0)}_\pm+\delta
n^{(1)}_\pm$ where
\begin{subequations}
\begin{equation}
\delta
n^{(0)}_\pm\!\!=\!-\frac{1}{\Pi_\pm}\left[\frac{iuq^2}{\sqrt{m c_v
T}}(\mathbf{qg}_\pm)
+(\omega_\chi-i\omega)\frac{\mathbf{q}}{m}(\hat{\varsigma}_\pm\mathbf{q})\right],
\end{equation}
\begin{equation}
\delta n^{(1)}_\pm\!\!=\!\frac{i(\mathbf{qv})}{2\Pi_\pm}
\left[\Gamma_+\delta n^{(0)}_++\Gamma_-\delta n^{(0)}_--
\frac{\mathbf{q}}{m}(\hat{\varsigma}_+\mathbf{q})-
\frac{\mathbf{q}}{m}(\hat{\varsigma}_-\mathbf{q})\right].
\end{equation}
\end{subequations}

\begin{figure}
  \includegraphics[width=8cm]{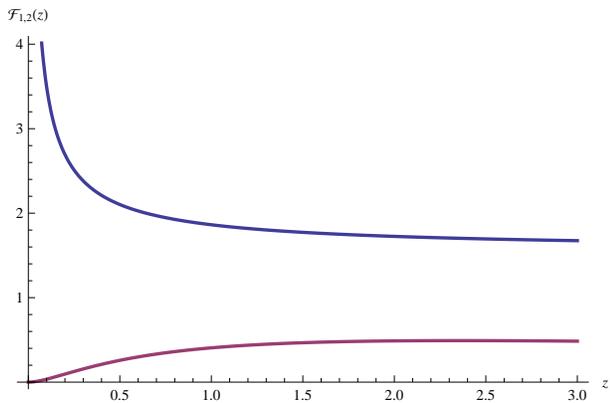}\\
  \caption{Dimensionless functions $\mathcal{F}_1$ (bottom line) and $\mathcal{F}_2$
  (top line)
  that enter the drag resistivity are plotted
  versus their respective scaling variable $z=\alpha$ and $z=\beta$,
  see Eqs.~\eqref{R-D}-\eqref{F-2} for the definition.}\label{Fig}
\end{figure}

Having found $\delta n$ we are in a position to compute the
density-density correlation function that determines the drag force
in Eq.~\eqref{F}. For this purpose we use thermal averages
$\langle\mathbf{q}(\hat{\varsigma}_\pm\mathbf{q})
\mathbf{q}(\hat{\varsigma}_\pm\mathbf{q})\rangle=4Tq^4(\eta+\zeta)$
and
$\langle(\mathbf{q}\mathbf{g}_\pm)(\mathbf{q}\mathbf{g}_\pm)\rangle
=4\kappa q^2T^2$, which follow directly from the Langevin heat flux
variances, and find
\begin{eqnarray}\label{rho-rho}
\langle\delta n_\uparrow(\mathbf{q},\omega)\delta
n_\downarrow(-\mathbf{q},-\omega)\rangle=
2in(\mathbf{q}\mathbf{v})\frac{q^2}{m}T(\omega^2_+-\omega^2_-)\nonumber\\
\times\frac{\omega^2_\nu\omega^4+2\omega_\nu\omega_\chi(\omega_\nu\omega_\chi+2\omega^2_\alpha)\omega^2
+\omega^2_\chi(\omega^2_\nu\omega^2_\chi-\omega^4_\alpha)}{|\Pi_+|^2|\Pi_-|^2}.
\end{eqnarray}
Before performing frequency integration in Eq.~\eqref{F} with the
density correlator from Eq.~\eqref{rho-rho} we need to analyze the
structure of poles of the propagator $\Pi_\pm$. Under the physically
relevant simplifying condition
$\omega_{\nu,\chi}\ll\mathrm{min}\{\omega_\alpha,\omega_\pm\}$,
which is justified by the fact that typical momentum transferred
between the layers is small $q\sim d^{-1}\ll\sqrt{n}$, we see that
$\Pi_\pm|_{\omega_\chi=0}=-i(\omega+i0)(\omega^2+i\omega\omega_\nu-\omega^2_\pm)$.
Already at this level we can identify plasmon poles at energies
$\omega_\pm$ whose imaginary part (lifetime) is governed by the
diffusive viscous mode $\omega_\nu$. Since $\omega_\nu\propto q^2$
fluctuations with sufficiently low momenta have arbitrary large mean
free path, and therefore plasmons are well defined excitations. At
the finite but small $\omega_\chi$ we can identify another pole,
$\omega+i0\to\omega+i\omega_\chi(1+\omega^2_\alpha/\omega^2_\pm)$,
which is governed by the thermal diffusion mode. We conclude that
density fluctuations that belong to one of the two parametrically
distinct frequency ranges $\omega\sim\omega_\chi$ and
$\omega\sim\omega_\pm$ give the largest contribution to the drag
force. Integrating Eq.~\eqref{F} within the leading pole
approximation we find our main result for the drag resistivity
\begin{equation}\label{R-D}
\rho_D=\frac{1}{16\pi^2e^2}\frac{1}{nd^2}\left[\frac{T}{\chi n
}\mathcal{F}_1(\alpha)+\frac{2\pi \nu T}{
\varpi^2nd^4}\mathcal{F}_2(\beta)\right],
\end{equation}
where $\varpi=\sqrt{2\pi e^2n/\epsilon md}$ is the plasmon energy
taken at the wave vector corresponding to the interlayer separation.
Two dimensionless parameters here are
\begin{equation}
\alpha=\left(\frac{u}{\varpi d}\right)^2,\qquad
\beta=\left(\frac{\nu}{\varpi d^2}\right)^2,
\end{equation}
and the two respective dimensionless functions are defined by the
following momentum integrals in the rescaled units
\begin{eqnarray}
&&\hskip-.65cm \mathcal{F}_1(\alpha)=\int^{\infty}_{0}\!\!
\frac{4\pi \alpha^2x^3e^{-2x}\mathrm{d}x}
{[(\alpha x+1)^2-e^{-2x}][(\alpha x+1)-e^{-2x}]},\label{F-1}\\
&&\hskip-.65cm \mathcal{F}_2(\beta)=\int^{\infty}_{0}\!\!
\frac{2x^4(\beta x^3+1)e^{-2x}\mathrm{d}x}{[1-e^{-2x}][\beta
x^3+e^{-2x}]}.\label{F-2}
\end{eqnarray}
When deriving above expressions we used
$(\omega^2_+-\omega^2_-)/\omega^2_+\omega^2_-=(\omega^2_p\sinh(qd))^{-1}$,
and rescaled momentum integration in the units of the interlayer
distance ($x=qd$). The first term in Eq.~\eqref{R-D} stems from the
slow thermal modes and is inversely proportional to the thermal
conductivity, while the second one is due to plasmons. We stress
that Eq.~\eqref{R-D} represents a nonperturbative in interaction
result for the drag resistivity and as such applies to electron
bilayers with $r_s\gg1$. The functions $\mathcal{F}_{1,2}$ are
plotted in Fig.~\eqref{Fig} and they are almost constants of the
order of unity $\mathcal{F}_{1,2}\sim1$ in a wide parameter range
$\{\alpha,\beta\}\gtrsim1$. For $\{\alpha,\beta\}\ll1$, which is
most likely relevant to experiments, one easily finds that
$\mathcal{F}_1\approx3\pi\zeta(3)\alpha^2/2$ and with the
logarithmic accuracy
$\mathcal{F}_2\approx\frac{1}{8}\ln^4(1/\beta)$, so that
Eq.~\eqref{R-D} can be simplified to
\begin{eqnarray}
\frac{\rho_D}{\rho_Q}\simeq\frac{3\varsigma[3]\epsilon^2T^3}{128\pi^3
e^4\kappa c_vn^3 d^4}\left(\frac{\partial s}{\partial\ln
n}\right)^4_T\nonumber\\+\frac{\epsilon
T(\eta+\zeta)}{128\pi^2e^2n^4d^5}\ln^{4}\left(\frac{2\pi
e^2mn^3d^3}{\epsilon(\eta+\zeta)^2}\right),\label{R-D-simple}
\end{eqnarray}
where $\varsigma[z]$ is the Riemann zeta function. Interestingly,
thermal expansion and plasmon mediated contributions to the Coulomb
drag resistivity Eq.~\eqref{R-D-simple} have distinct dependencies
on the electron density $n$ and interlayer separation $d$. One
should also notice that complete temperature dependence of
$\rho_D(T)$ is implicit in the corresponding temperature
dependencies of the respective kinetic and thermodynamic
coefficients $\kappa(T)$, $\eta(T)$, $\zeta(T)$ and $c_v(T)$.
Although a detailed microscopic theory for the temperature
dependence of $\kappa$, $\eta$ and $\zeta$ of nondegenerate strongly
correlated liquids has not been developed, some conjectures were put
forward in Refs.~[\onlinecite{AFAndreev,Kivelson}]. In particular
for the semiquantum regime at $E_F<T<\Omega_D$ one estimates
$c_v\propto T$, $\kappa\propto T$, and $\eta\propto 1/T$.

In contrast, the Fermi-liquid regime has been studied
extensively,~\cite{Abrikosov} and for $T\lesssim E_F$ one readily
finds $\kappa\sim c_vn\ell_{\mathrm{ee}}v_F\sim E^2_F/T$, $\eta\sim
mv_Fn\ell_{\mathrm{ee}}\sim nE^2_F/T^2$, and $c_v\sim T/E_F$. Note
that for Fermi liquids the temperature dependence of the drag
resistivity in the hydrodynamic regime is drastically different from
the conventional $T^2$ law. Indeed, assuming that interaction
parameter $r_s\sim1$ one estimates the first term in
Eq.~\eqref{R-D-simple}  as $\sim (T/E_F)^7(1/k_Fd)^4$ and the second
one as $\sim (E_F/T)(1/k_Fd)^5$. At $T< E_F/(k_F d)^{1/8}$ the
second term dominates and we obtain the following estimate for the
drag coefficient of Fermi liquids in the hydrodynamic regime
\begin{equation}
\frac{\rho_D}{\rho_Q} \sim \frac{E_F}{T}\frac{1}{(k_Fd)^5}.
\end{equation}
The hydrodynamic description is restricted to temperatures
$T>E_F/\sqrt{k_Fd}$ where $\ell_{\mathrm{ee}}<d$. It is striking to
observe that at temperatures where $\ell_{\mathrm{ee}}\sim d$ the
hydrodynamic result above, $\rho_D/\rho_Q\sim(1/k_Fd)^{9/2}$, is
parametrically larger (in $\sqrt{k_Fd}\gg1$) than the conventional
FL result for the collisionless regime,
$\rho_D/\rho_Q\simeq(T/E_F)^2/(k_Fd)^4\sim(1/k_Fd)^5$. This implies
that collisions strongly enhance Coulomb drag. The study of the
crossover from a collisionless to a collision-dominated regime is an
interesting problem that is beyond the scope of the present work.

\textit{Discussion}. It is perhaps instructive to compare our result
for the drag resistivity Eq.~\eqref{R-D-simple} to the hydrodynamic
result for intralayer resistivity in Ref.~\onlinecite{Andreev} [see
their Eq.~(6)]. Both are given by the sum of thermal and viscous
terms, which have similar dependence on the viscosity and thermal
conductivity of the fluid. This is not accidental. To second order
in the disorder potential the intralayer resistivity can be
understood in terms of the drag force between the electron liquid
and the disorder potential $\mathbf{F}_D=\langle\delta n(-\nabla
U)\rangle$ with $U$ representing the disorder potential. In that
case, the fluctuations of density in the electron liquid are created
by the disorder potential itself. The subsequent scattering of
density fluctuations from the disorder potential produces a net
resistive force. In the case of drag, both the scattering potential
$U$ and the fluctuations of the electron density are produced by
thermal fluctuations, whose variance depends on the temperature.
This accounts for the difference between the temperature dependence
of drag and intralayer resistivity. On the other hand, the
propagation of fluctuations in the fluid in either case is described
by the same linearized hydrodynamic equations, and occurs in the
form of stress-driven ballistic modes and entropy-driven diffusive
modes. This results in the similarity between the corresponding
expressions.

The above qualitative arguments are useful in contrasting our theory
with the energy transfer mechanism of Coulomb drag studied in the
context of graphene double layers.~\cite{Levitov} The energy
transfer mechanism ($E$-drag) does not involve thermal fluctuations,
and may be treated in the main hydrodynamic approximation. On the
other hand, this mechanism relies on correlations of the disorder
potential in the layers and disappears in the clean limit. The
contribution considered in the present paper arises from fluctuation
corrections to hydrodynamics, and remains finite in the clean limit.
The common feature of two mechanisms is that diffusive
redistribution of thermal energy in the electron liquid plays a
crucial role in supporting strong Coulomb drag.

\textit{Acknowledgments}. We would like to thank B.~Spivak for many
useful discussions, and J.~Song and L.~Levitov for discussions on
the energy-driven mechanism of Coulomb drag. A.L. acknowledges
support from NSF under Grant No. PHYS-1066293 and the hospitality of
the Aspen Center for Physics where part of this work was performed.
The work of A. V. A. was supported by DOE Grant No.
DE-FG02-07ER46452.


\begin{thebibliography}{99}
\bibitem{Note}
To be clear, we consider a large $r_s$ limit but still assume its
value to be smaller than a critical one $r_s\lesssim
r^{(c)}_{s}\sim40$ when electron fluid crystallizes into the Wigner
state~\cite{Tanatar}. At a finite temperature the condition is less
stringent because the Wigner crystal may be melted by thermal
fluctuations.
\end{thebibliography}
\end{document}